\documentclass[12pt]{article}
\usepackage[dvips]{color}
\usepackage{epsfig}
\usepackage{amsmath}
\usepackage{graphicx}

\begin{document}
\title{\bf PV criticality of the second order quantum corrected Ho\v{r}ava-Lifshitz black hole}
\author{{ B. Pourhassan$^{a}$\thanks{Email: b.pourhassan@du.ac.ir}}\\
$^{a}${\small {\em School of Physics, Damghan University, P. O. Box 3671641167, Damghan, Iran}}}
\maketitle
\begin{abstract}
\noindent In this paper, higher order quantum gravity effects on the thermodynamics of Ho\v{r}ava-Lifshitz black hole investigated. Both Kehagius-Sfetsos and Lu-Mei-Pop solutions of Ho\v{r}ava-Lifshitz black hole considered and higher order corrected thermodynamics quantities obtained. The first order correction is logarithmic and second order correction considered proportional to the inverse of entropy. These corrections are due to the thermal fluctuation and interpreted as quantum loop corrections. Effect of such quantum corrections on the stability and critical points of Horava-Lifshitz black holes studied. We find that higher order correction affects critical point and stability of Lu-Mei-Pop solution and yield to the second order phase transition for the case of Kehagius-Sfetsos solution.
\end{abstract}
\section{Introduction}
Black holes are gravitational objects with maximum entropy \cite{1, 2} which is related to the black hole event horizon area $A$. In that case, statistical mechanics help us to study the black hole thermodynamics and it is one of the important field of theoretical physics \cite{wald, sorkin}. Hence, the thermodynamics study of several kinds of black hole is important subject of the recent researches. Lower dimensional black holes are interesting and usually considered as a toy model to obtain gravitational solutions. For example, two-dimensional black holes may be considered to study thermodynamics \cite{2d1,2d2,2d3,2d4,2d5,2d6}. Higher dimensional black holes are also interesting in several theories like near-extremal solutions of Einstein-Maxwell-scalar theory \cite{EMs}, or hyperscaling violation backgrounds \cite{hyper1, hyper2, hyper3}.\\
Due to the thermal fluctuations, the black hole entropy modified and correction terms interpreted as quantum effect, because the quantum gravity modified the manifold structure of space-time at Planck scale \cite{6b, 6a}. Almost all methods of quantum gravity indicated that the leading order correction is logarithmic \cite{l1, SPR} and it has been argued that the general structure of the correction terms is a universal. In that case, leading order quantum corrections to the geometry of large AdS black holes and their effects on the thermodynamics given by the Ref. \cite{th}. Hence, the thermal fluctuations in a hyperscaling violation background considered by the Ref. \cite{hyper4}. Thermodynamics of Kerr-Newman-AdS black holes already studied by the Ref. \cite{CCK}.
Similar black hole (uncharged but $d$-dimensional) considered under the effect of the thermal fluctuations \cite{NPB} and found that logarithmic correction becomes important when the size of the black hole becomes small due to the Hawking radiation \cite{page}. Also, statistical mechanics of charged black holes considered by the Ref. \cite{Leb}. Black hole thermodynamics in modified theories of gravity like $f(r)$ \cite{f,ref6,ff,fff} also studied by the Ref. \cite{mir}.\\
It is also possible to obtain higher order corrections \cite{more, CJP} and such corrected terms has the same universal shape as expected from the quantum gravitational effects \cite{l1, SPR, more}.  Black holes in G\"{o}del universes already introduced by the Ref. \cite{Gimon}, in that case Kerr-G\"{o}del black hole thermodynamics investigated by the Ref. \cite{godel1}. Logarithmic correction to the G\"{o}del black hole has been studied by Ref. \cite{godel}. Higher order correction to the Kerr-Newman-G\"{o}del black hole recently given by the  Ref. \cite{godel2} and demonstrated that second order correction is proportional to the inverse of entropy. Thermodynamics of higher order entropy corrected Schwarzschild-Beltrami-de Sitter black hole investigated by the Ref. \cite{ref7} and found that higher order corrections may affect the black hole stability.\\
STU black holes \cite{STU2} are other interesting kinds of black hole which has been studied by the Ref. \cite{STU1} from statistical point of view \cite{car}. This kind of black hole is interesting in AdS/CFT correspondence \cite{Ross}, for example it is possible to compute hydrodynamics and thermodynamics properties of quark-gluon plasma \cite{hydro1,hydro2, hydro3, hydro4} in presence of quantum correction. Other properties of quark-gluon plasma like drag force \cite{drag1,drag2,drag3,drag4}, jet-quenching \cite{jet1,jet2,jet3,jet4} and shear viscosity to entropy ratio \cite{EPJC} may also affected by thermal fluctuations, and this can observed by experiments \cite{QGPexp}.\\
It is important to note that corrections to the black hole thermodynamics can study using the non-perturbative quantum  general relativity. Moreover, it is possible to investigate black hole thermodynamics under effects of thermal fluctuations by using the effect of matter fields surrounding a black hole \cite{other, other0, other1}. The corrected thermodynamics of a dilatonic black hole from Cardy formula have been already studied by the Ref. \cite{jy} and show the same universal form of correction term as previous studies. This universality can be understood in the Jacobson formalism  \cite{z12j, jz12}. Corrected thermodynamics of a black hole can be obtained by using the partition function \cite{bss}.
Such corrections have already considered for different black objects. For example, the first order correction of the AdS charged black hole has been studied, and modified thermodynamics obtained \cite{1503.07418}, which is extended to the case of AdS charged rotating black hole \cite{ref3}. In the Ref. \cite{1503.07418} it is argued that leading order correction is logarithmic its coefficient can be considered as free parameter of the theory.  Thermal fluctuations effect of a black saturn have been studied \cite{1505.02373}.
It was observed by considering charged dilatonic black saturn that the thermal fluctuations can be obtained either using a conformal field theory or by analyzing the fluctuations in the energy of this system and both yields to the similar results for a charged dilatonic black saturn \cite{1605.00924}.\\
An interesting kind of regular black holes is Hayward black hole \cite{Hayward} which can be modified \cite{Hayward1, Hayward2}. In that case, the logarithmic corrections of a modified Hayward black hole considered to calculate some thermodynamics quantities, and found that
this leading order correction affect the pressure and internal energy by decreasing value of them \cite{1603.01457}. Thermal fluctuations of charged black holes in gravity's rainbow investigated by the Ref. \cite{ref4}, and quantum corrections to thermodynamics of quasitopological black holes studied by the Ref. \cite{ref5}.\\
Investigation of thermal fluctuation in gravitational systems may help us to test the quantum gravity effects, for examples, on
dumb holes \cite{Annals} or graphene \cite{Graphene}. Logarithmic and higher order corrections may affect the critical behaviors of black objects, for example in the Ref. \cite{PRD} a dyonic charged anti-de Sitter black hole thermodynamics considered and show that holographic picture (a van der Waals fluid) is still valid. In that case logarithmic corrected van der Waals black holes in higher dimensional AdS space investigated by the Ref. \cite{PTEP}. Also, thermodynamics of higher dimensional black holes with higher order thermal fluctuations studied in \cite{GRG}.\\
BTZ black holes \cite{ref1} also considered to investigate effects of thermal fluctuations \cite{ref2} including higher order corrections \cite{BTZ1, BTZ2, BTZ3}. P-V criticality of black holes also may affected by thermal fluctuations \cite{PV, PV1, PV2}.\\
Ho\v{r}ava-Lifshitz (HL) black holes are important kind of black holes in theoretical physics \cite{4}. The HL gravity is also an interesting theory of quantum gravity \cite{5,6,7,8} which considered in particle physics and cosmological literatures \cite{dark, CJP2}. We expect that the HL black hole
solutions, asymptotically, become Einstein gravity solutions. In that case, Refs. \cite{main, 9, 10, 11} investigated thermodynamics quantities of HL black holes. It has been reported some instabilities in HL black holes. Hence, such thermodynamics using logarithmic corrected entropy investigated to find quantum gravity effects \cite{NPB2} and found that some instabilities removed due to thermal fluctuations.\\
In the Ref. \cite{NPB2}, only LMP solutions of HL black hole considered and first order correction (logarithmic correction) investigated. Now, in this paper, we would like to consider both LMP and KS solutions and investigate effects of higher order quantum corrections.\\
This paper organized as follows. In the section 2 we review higher order correction and propose a general form with free correction coefficients. In section 3, Ho\v{r}ava-Lifshitz (HL) black holes properties called. General thermodynamics relations introduced in section 4. In section 5, effects of higher order corrections on the thermodynamics of Lu-Mei-Pop solution investigated for three different cases of flat, spherical and hyperbolic spaces. In section 6 we consider Kehagius-Sfetsos solution of HL black hole and study corrected thermodynamics.
Finally, in section 7 we give conclusion.

\section{Higher order corrections}
In the canonical ensemble, one can calculate the density of state. Partition function of $N$ particles given by the following expression \cite{hawk, hawk1},
\begin{equation}\label{Z}
Z=\int_{0}^{\infty}\rho(E)e^{-\beta E}dE,
\end{equation}
where $E$ is the average energy in the canonical ensemble. Having partition function, one can write entropy as,
\begin{equation}\label{S beta}
S(\beta)=\beta E+\ln{Z}.
\end{equation}
Now, it is possible to consider thermal fluctuations and use Taylor expansion around $\beta$ and write,
\begin{equation}\label{S}
S(\beta)=S_{0}(E)+\frac{(\beta-\beta_{0})^{2}}{2}\left(\frac{\partial^{2}}{\partial\beta^{2}}\ln{Z}\right)_{\beta_{0}}
+\frac{(\beta-\beta_{0})^{3}}{3!}\left(\frac{\partial^{3}}{\partial\beta^{3}}\ln{Z}\right)_{\beta_{0}} +\cdots,
\end{equation}
where maximum of entropy at equilibrium point $\beta=\beta_{0}$ established. It can written as,
\begin{equation}\label{fluc}
S = S_0 + \frac{1}{2}(\beta - \beta_0)^2 \left(\frac{\partial^2 S(\beta)}{\partial \beta^2 }\right)_{\beta = \beta_0},
\end{equation}
where $S_0$ is uncorrected entropy.\\
Indeed, correction terms are due to statistical fluctuations around equilibrium, hence one can assume the system in equilibrium and use usual thermodynamics relations. Equation (\ref{S}) yields to a general form for different black geometries,
\begin{equation}\label{CS}
S=S_{0}+\alpha\ln{S_{0}}+\frac{\gamma}{S_{0}}+\cdots.
\end{equation}
In this paper, we would like to study the second order correction hence neglect higher order terms. $\alpha$ and $\gamma$ consider as free parameter of the theory and they are depending on black hole physics. For example it is found that $\alpha=-\frac{3}{2}$ and $\gamma=-\frac{3}{16}$ for the BTZ black hole, $\alpha=-1$ and $\gamma=-\frac{5}{36}$ for the AdS Schwarzschild black hole \cite{more}. Now, we can examine some values of free parameters to see their effects on the thermodynamics quantities of HL black hole. Indeed we would like to test only presence of corrections with positive or negative values. Calculation of of exact values of $\alpha$ and $\gamma$ are also interesting which is not aim of this paper. In the next section, we give brief review of Ho\v{r}ava-Lifshitz black holes.

\section{Ho\v{r}ava-Lifshitz Black Hole}
First of all, we review some important properties of HL black holes.
The action of HL theory in four-dimensional gravity is given by \cite{12,13,14,15,16},\\
\begin{eqnarray}\label{1}
S_{HL}&=&\int dtd^{3}x\sqrt{g}N\left(\frac{2}{\kappa^2}(K_{ij} K^{ij}-\lambda K^{2})+\frac{\kappa^2 \mu^2(\Lambda_W
R-3\Lambda_W^2)}{8(1-3\lambda)}+\frac{\kappa^2\mu^2(1-4\lambda)}{32(1-3\lambda)}R^{2}\right)\nonumber\\
&-&\int dtd^{3}x\sqrt{g}N\left(\frac{\kappa^2\mu^2}{8}R_{ij}R^{ij}
 +\frac{\kappa^2\mu}{2\omega^2}\epsilon^{ijk} R_{il}\nabla_j
R_k^{l} - \frac{\kappa^2}{2\omega^4}C_{ij}C^{ij}\right),
\end{eqnarray}
where $\kappa^2$, $\lambda$, $\omega$, $\Lambda$ and $\mu$ are constant parameters, and $C_{ij}$ is cotton tensor given by \cite{17-1,17-2},
\begin{equation}\label{2}
C^{ij}=\epsilon^{ikl} \nabla_k (R_{l}^{j} - \frac{1}{4} R\delta_{l}^{j}),
\end{equation}
and $K_{ij}$ is extrinsic curvature,
\begin{equation}\label{3}
K_{ij}=\frac{1}{2N}(\dot{g}_{ij} -\nabla_i N_j - \nabla_j N_i ),
\end{equation}
where $N_i$ is shift function and $N$ is lapse function, also the cosmological constant is given by $\Lambda=\frac{3}{2}\Lambda_{W}$ \cite{18}, where $\Lambda_{W}$ is a constant negative parameter (we consider $\Lambda_{W}=-l$).\\
The corresponding metric is given by \cite{12},
\begin{equation}\label{5}
ds^2=f(r)dt^2 - \frac{dr^2}{f(r)} - r^{2}d\Omega_{2}^{2},
\end{equation}
where,
\begin{eqnarray}\label{6}
d\Omega_{2}^{2}\equiv\big\{\begin{array}{ccc}
d\theta^2 + \sin^{2}\theta d\varphi^2 \hspace{1.2cm} (k=1)\\
d\theta^2 + \theta^2 d\varphi^2\hspace{1.8cm}(k=0)\\
d\theta^2 + \sinh^{2}\theta d\varphi^2\hspace{1.2cm}(k=-1)\\
\end{array}
\end{eqnarray}
and $f(r)$ is given by the following relation \cite{13},
\begin{equation}\label{7}
f(r)=k + (\omega - \Lambda_W)r^{2} -\sqrt{(r(\omega(\omega - 2\Lambda_W)r^{3} + B))},
\end{equation}
where $B$ is an integration constant.\\
In the case of $\Lambda_W=0$ and $B=4\omega M$ we have Kehagius-Sfetsos (KS) solution \cite{19},
\begin{equation}\label{8}
f_{ks} (r)=k +\omega r^{2}- \omega r^{2}\sqrt{1+\frac{4M}{\omega r^{3}}},
\end{equation}
while in the case of $\omega=0$ and $B=-\frac{\alpha^2}{\Lambda_W}$ we have Lu-Mei-Pop (LMP) solution \cite{12},
\begin{equation}\label{9}
f_{LMP}(r)=k - \Lambda_Wr^{2} - aM \sqrt{\frac{r}{-\Lambda_W}}.
\end{equation}
Black hole thermodynamics in KS solution of HL Gravity have been studied by the Ref. \cite{18}, and thermodynamical quantities of
LMP solution of HL black hole for three different cases of spherical, flat and hyperbolic spaces given by the Ref. \cite{main}. Also, first order correction on LMP solution of HL black hole studied by the Ref. \cite{NPB2}. In the next section we write some important thermodynamics relations of HL black holes.
\section{Thermodynamics}
The first law of thermodynamic may be given by \cite{20}
\begin{equation}\label{12}
dM=TdS+VdP.
\end{equation}
The black hole entropy and Hawking temperature obtained by using the following equations \cite{21},
\begin{eqnarray}\label{13}
S_{0}&=&\int \frac{dM}{T}=\int \frac{1}{T_H}\frac{\partial H}{\partial
r_h}dr_h,\nonumber\\
T_{H}&=&\frac{\kappa}{2\pi}=\frac{1}{4\pi}(\frac{\partial
f}{\partial r})_{r=r_h},
\end{eqnarray}
where $H$ denotes the enthalpy and interpreted as the black hoe mass ($H=M$) \cite{22}. Also temperature of black hole is given by the following relation \cite{22},
\begin{equation}\label{14}
T=(\frac{\partial H}{\partial S_{0}})_P=(\frac{\partial H}{\partial
r_h}\frac{\partial r_h}{\partial S_{0}})_{P}.
\end{equation}
It is easy to check that $T=T_{H}$ and black hole pressure obtained by $P=\frac{1}{2}TS$ \cite{23,24}.
The volume of black hole is given by,
\begin{equation}\label{17}
V=(\frac{\partial H}{\partial P})=(\frac{\partial H}{\partial
r_h}) (\frac{\partial P}{\partial r_h})^{-1}.
\end{equation}
The heat capacity can be found by using the following relation \cite{22, 25},
\begin{equation}\label{19}
C=(\frac{\partial E}{\partial T})=T(\frac{\partial S}{\partial T}),
\end{equation}
where the internal energy is given by,
\begin{equation}\label{20}
E=H-PV=F+TS,
\end{equation}
where
\begin{equation}\label{21}
F=-\int{SdT},
\end{equation}
is Helmholtz free energy which related to the Gibbs free energy via,
\begin{equation}\label{21-1}
G = F + PV = H - TS.
\end{equation}
Thermodynamical quantities of the HL black hole for LMP and KS solutions are different for spherical space $(k=1)$, flat space $(k=0)$, and hyperbolic space $(k=-1)$.

\section{Lu-Mei-Pop solution}
LMP solution given by the equation (\ref{9}) and there are three different cases corresponding to space curvature.

\subsection{Spherical space}
In this case the black hole mass given by,
\begin{equation}\label{22}
M=a^{-1}\sqrt{\frac{-\Lambda_W}{r_h}}(1 -\Lambda_W r_h ^{2}).
\end{equation}
It has been found that the magnitude of the cosmological constant increases the black hole mass. However there is a minimum for the black hole
mass at $r_{h}=1/\sqrt{-3\Lambda_W}$.\\
By using the equation (\ref{13}) one can obtain the entropy and Hawking temperature as follow,
\begin{equation}\label{23}
S_{0}=\frac{8\pi}{a}\sqrt{-\Lambda_W r_h},
\end{equation}
and
\begin{equation}\label{24}
T=\frac{1}{8\pi r_h}[-1-3\Lambda_W r_h ^2].
\end{equation}
It has been also finding that the magnitude of the cosmological constant increases the black hole temperature. The zero temperature limit obtained by $r_{h}=1/\sqrt{-3\Lambda_W}$.\\
Now, we can use the equation (\ref{CS}) to obtain corrected thermodynamics due to higher order quantum corrections.\\
In this case the black hole Helmholtz free energy obtained using the relations (\ref{21}), (\ref{23}), (\ref{24}), and (\ref{CS}), as
\begin{eqnarray}\label{62}
F&=&2\sqrt {{\frac {l}{r_h}}}\frac{1-l{r_h}^{2}}{a}\nonumber\\
&+&{\frac {\alpha}{8\pi \,r_h} \left(  \left( 1-3\,l{r_h}^{2} \right) \ln  \left( 8\,{\frac {
\pi \sqrt {l r_h}}{a}} \right) +\frac{1}{2}(1+3l{r_h}^{2})\right) }\nonumber\\
&+&{\frac {a\gamma\,\sqrt {l}}{32{\pi }^{2}\sqrt {r_h}} \left({\frac {1
}{3l{r_h}^{2}}}-3 \right) },
\end{eqnarray}
where we set $\Lambda_{W}=-l$.\\

\begin{figure}[h!]
 \begin{center}$
 \begin{array}{cccc}
\includegraphics[width=60 mm]{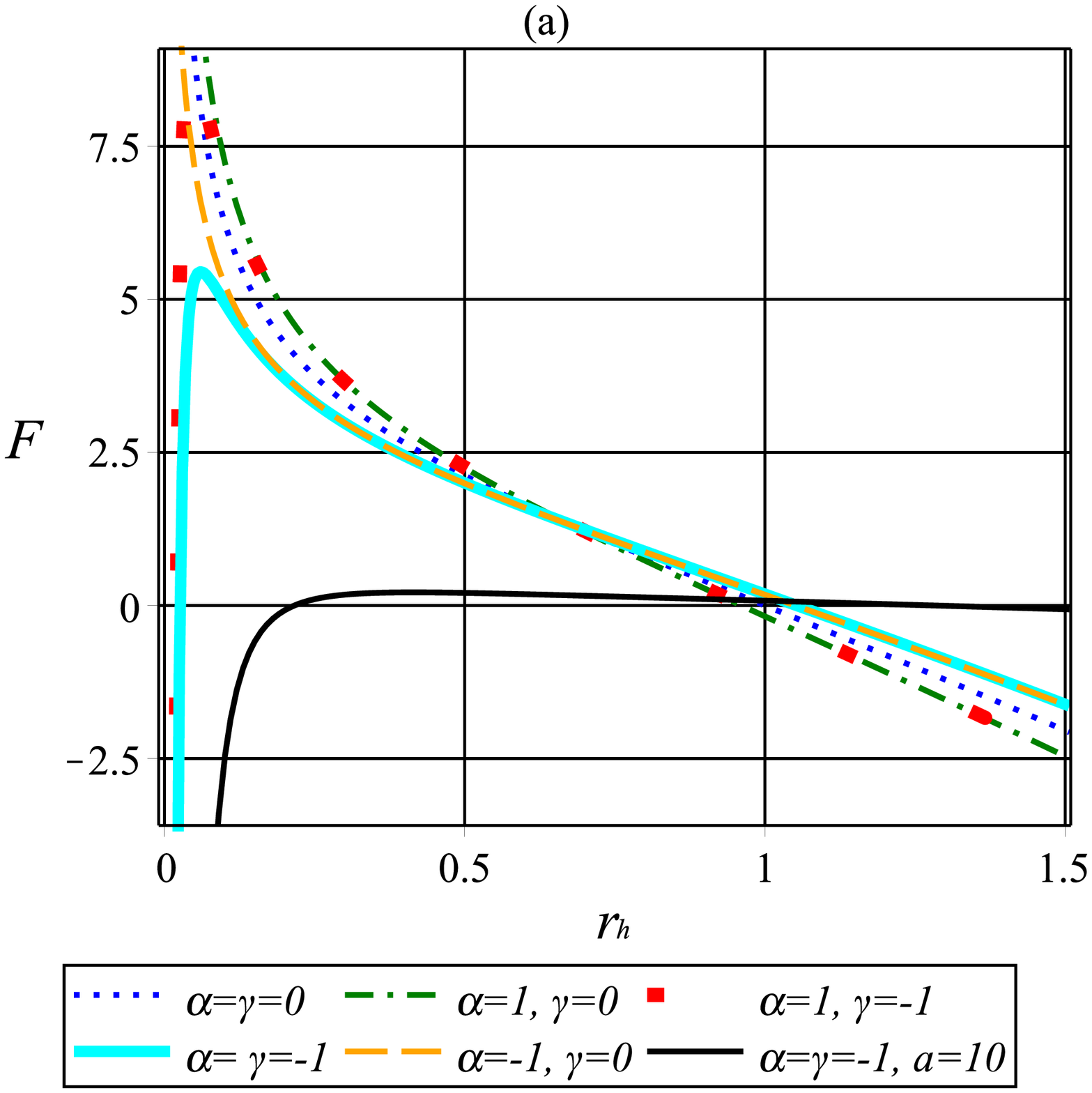}\includegraphics[width=60 mm]{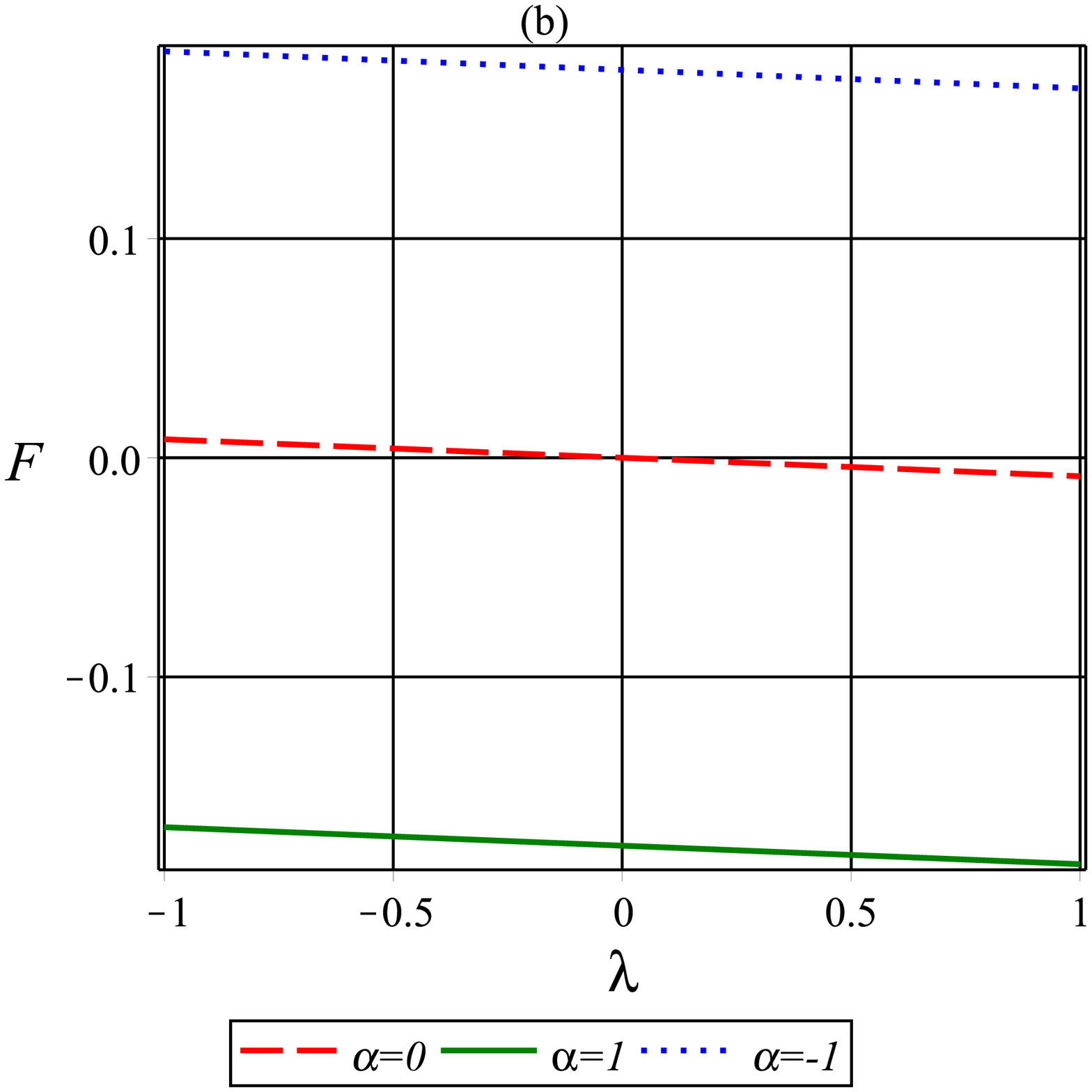}
 \end{array}$
 \end{center}
\caption{Helmholtz free energy of the LMP solution of HL black hole in spherical space for $l=a=1$. (a) in terms of horizon radius, (b) in terms of the second order correction parameter with $r_{h}=1$.}
 \label{fig1}
\end{figure}

\begin{figure}[h!]
 \begin{center}$
 \begin{array}{cccc}
\includegraphics[width=55 mm]{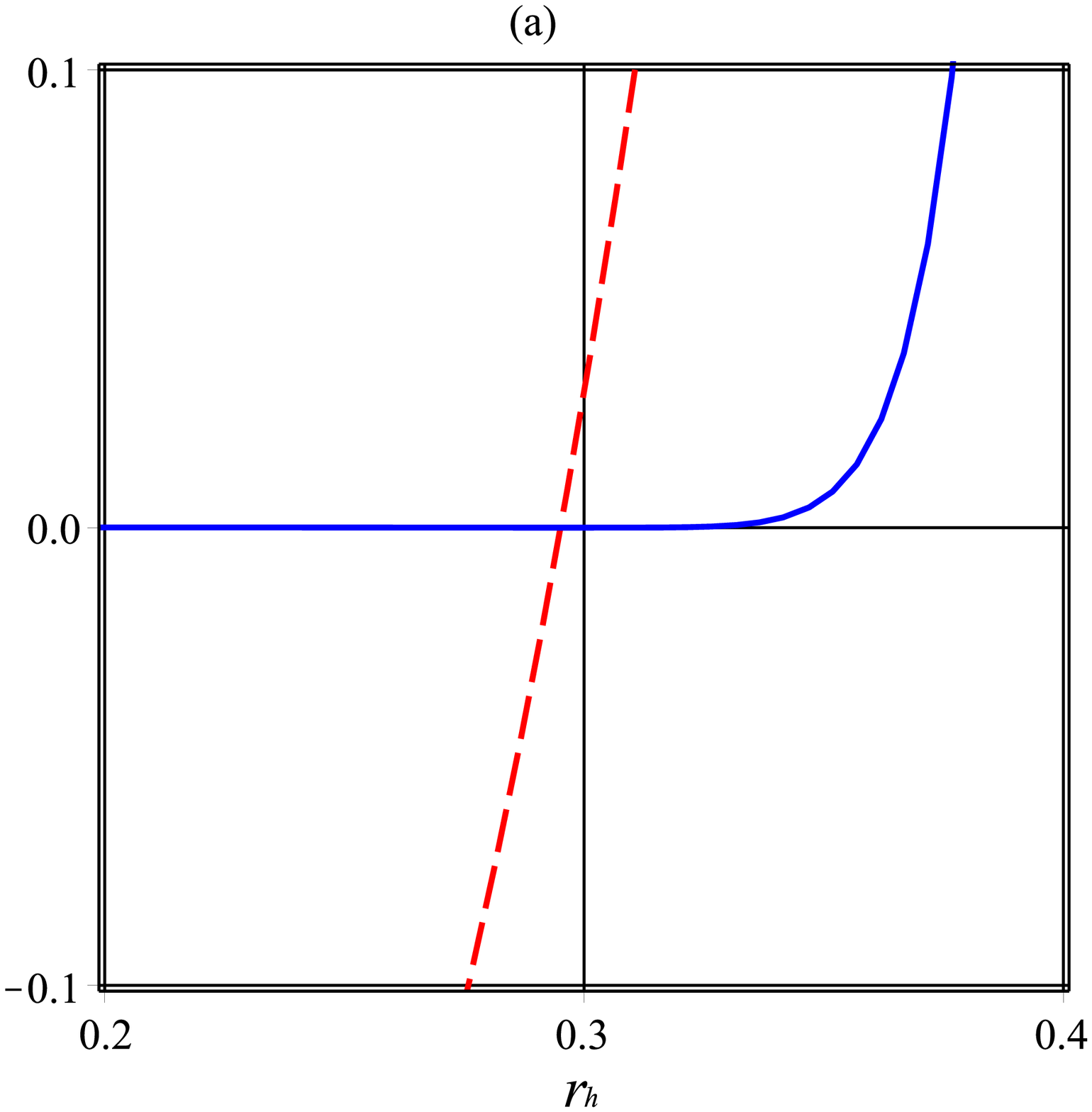}\includegraphics[width=55 mm]{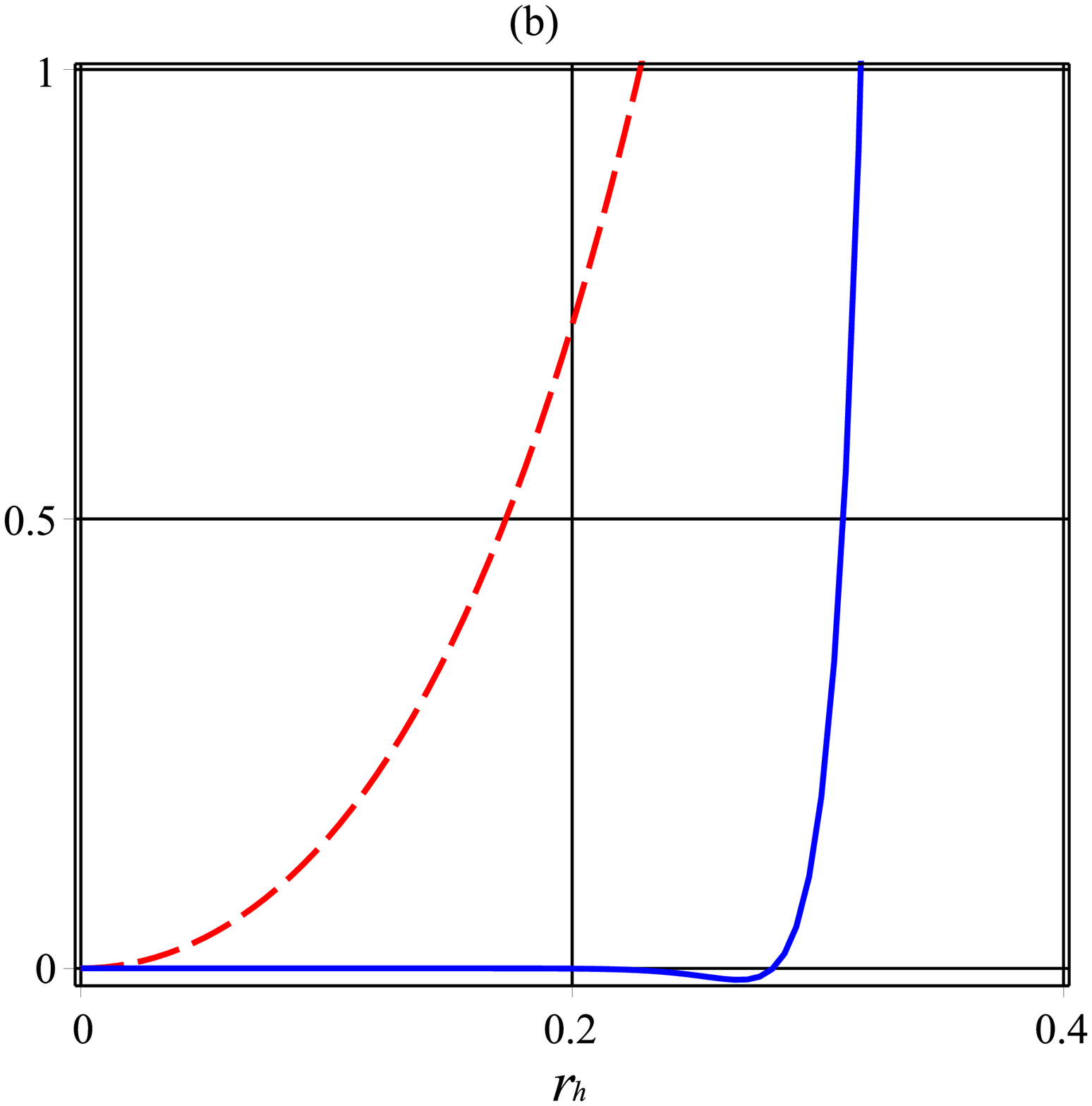}\includegraphics[width=55 mm]{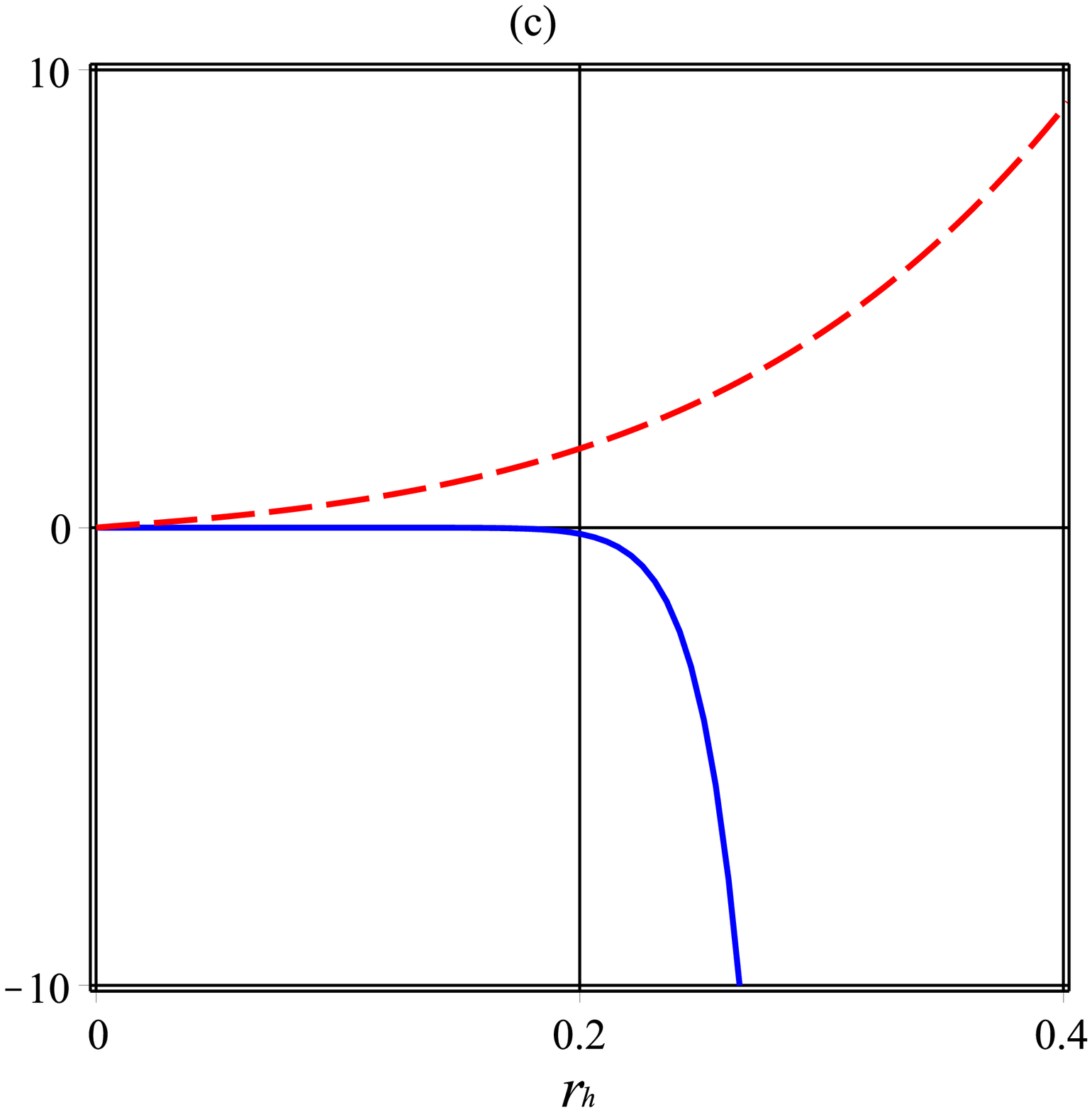}
 \end{array}$
 \end{center}
\caption{Critical points of LMP solution in spherical space for $l=1$ and $a=10$: (a) $\alpha=\gamma=-1$. (b) $\alpha=\gamma=0$. (c) $\alpha=\gamma=1$. Dashed red lines represent $\frac{dP}{dV}$, while solid blue lines represent $\frac{d^{2}P}{dV^{2}}$.}
 \label{fig2}
\end{figure}

In the Fig. \ref{fig1} we can see behavior of the Helmholtz free energy and obtain the effect of the quantum corrections. We examined both positive and negative values of correction coefficients. Generally, we find that Helmholtz free energy is decreasing function of correction parameters (see Fig. \ref{fig1} (b)). It means that quantum corrections reduced value of Helmholtz free energy. In the case of $\gamma=0$ we recover result of the Ref. \cite{NPB2}.
We can see that there is special radius where corrected and uncorrected curves crossed, which means thermal fluctuations have no any effect in special radius. In the Fig. \ref{fig1} this special radius is about $r_{h}=0.7$. There is a maximum value for the Helmholtz free energy at small radius, which we can see by solid cyan line of the Fig. \ref{fig1} (a) corresponding to the second order quantum correction with $\alpha=\gamma=-1$. Such maximum may relate to the critical point of the black hole.\\
One can obtain pressure as follow,
\begin{equation}\label{64}
P=\frac{3lr_{h}^{2}-1}{16\pi r_{h}}\left[\frac{8\pi}{a}\sqrt{lr_{h}}+\alpha\ln{\frac{8\pi\sqrt{lr_{h}}}{a}}+\frac{\gamma a}{8\pi \sqrt{lr_{h}}}\right],
\end{equation}
while the black hole volume by using the equation (\ref{17}) obtained as follow,
\begin{equation}\label{VLMPSP}
V=\frac{128(3lr_{h}^{2}-1)\pi^{2}l^{2}r_{h}^{2}}{64(9lr_{h}^{2}+1)\pi^{2}l^{2}r_{h}^{2}+8\pi a (lr_{h})^{\frac{3}{2}}\alpha(3lr_{h}^{2}+1)(1+2\ln{\frac{8\pi\sqrt{lr_{h}}}{a}})+3la^{2}r_{h}\gamma(lr_{h}^{2}+1)}.
\end{equation}
Having $P$ and $V$, we can study $P-V$ critical points where
\begin{equation}\label{CriticalCondition}
\frac{dP}{dV}=\frac{d^{2}P}{dV^{2}}=0.
\end{equation}
The points which satisfy condition (\ref{CriticalCondition}) called critical points. In the Fig. \ref{fig2} we analyze critical points for the ordinary case ($\alpha=\gamma=0$) and corrected cases (Figs. \ref{fig2} (a) and (c)). Dashed red lines represent $\frac{dP}{dV}$, while solid blue lines represent $\frac{d^{2}P}{dV^{2}}$, when both curves cross each other in zero axis, then the corresponding $r_{h}$ is the black hole radius where critical points happen. We can see from the Figs. \ref{fig2} (a) that only in the case of $\alpha=\gamma=-1$ (presence of higher order correction) we have critical points near $r_{h}\approx0.3$. It is corresponding to black solid line of the Fig. \ref{fig1} (a) at maximum point of Helmholtz free energy. It means that in order to have critical points, presence of the higher order corrections is necessary. For the cases of correction with positive coefficients and uncorrected entropy the critical points exist only at $r_{h}=0$.\\

Heat capacity is important parameter to study stability of the black hole. By using the equation (\ref{19}), we can calculate heat capacity as,
\begin{equation}\label{29}
C=\frac{3lr_{h}^{2}-1}{16\pi a(lr_{h})^{\frac{3}{2}}(3lr_{h}^{2}+1)}\left(64\pi^{2}l^{2}r_{h}^{2}+8\pi a (lr_{h})^{\frac{3}{2}}\alpha-la^{2}r_{h}\gamma\right).
\end{equation}
It is clear that logarithmic correction increases value of heat capacity while the second order correction reduces its value.\\

\begin{figure}[h!]
 \begin{center}$
 \begin{array}{cccc}
\includegraphics[width=75 mm]{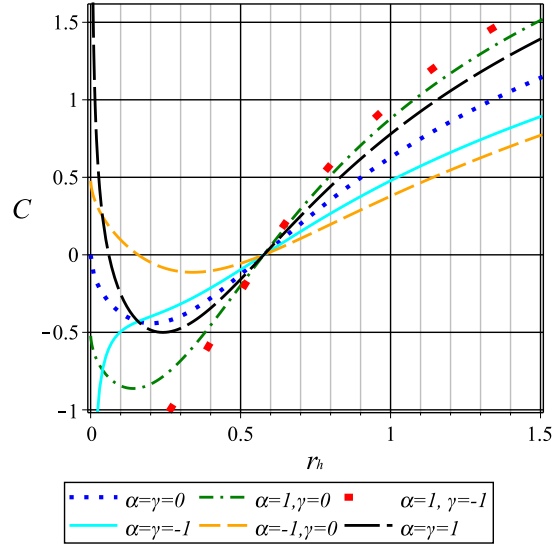}
 \end{array}$
 \end{center}
\caption{Specific heat of LMP solution in spherical space in terms of the black hole horizon for $l=1$ and $a=10$.}
 \label{fig3}
\end{figure}

\begin{figure}[h!]
 \begin{center}$
 \begin{array}{cccc}
\includegraphics[width=75 mm]{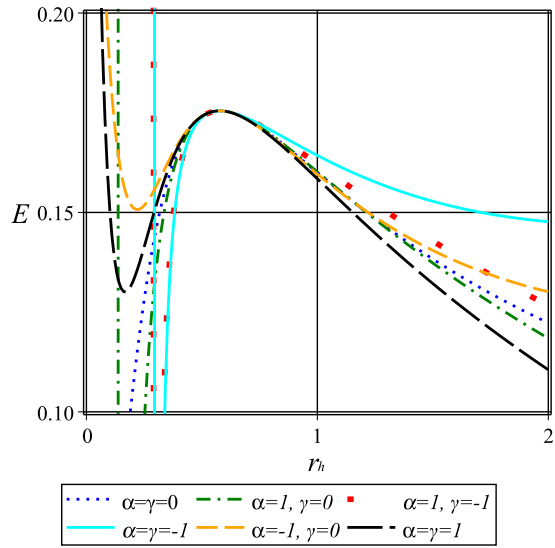}
 \end{array}$
 \end{center}
\caption{Internal energy of LMP solution in spherical space in terms of the black hole horizon for $l=1$ and $a=10$.}
 \label{fig4}
\end{figure}

Without the logarithmic correction we saw some instabilities corresponding to the small horizon radius (see dotted blue line of Fig. \ref{fig3}). It has been found that the magnitude of the cosmological constant increases the black hole heat capacity. The zero heat capacity limit obtained by $r_{h}=1/\sqrt{3l}$. It means that there are some instabilities for small $r_{h}$ where $C$ is negative. Although, quantum corrections have not any important effect on this point but may make other stable points at very small radius. In all cases there is no any asymptotic behavior corresponding to phase transition. Solid cyan line of the Fig. \ref{fig3} is corresponding to the case of $\alpha=\gamma=-1$ which shows a turning point near $r_{h}\approx0.3$ (critical point and maximum of Helmholtz free energy). Some cases of corrected heat capacity (dashed lines) show that there is some stable region for small radius, then the black hole is unstable in a finite rage of $r_{h}$. It means that, when the black hole size reduced due to the Hawking radiation, and thermal fluctuations become important, then the black hole is stable at quantum scales. Hence, we can't neglect thermal fluctuations of small black holes.\\
Finally the internal energy is obtained using the equation (\ref{20}). We give graphical analysis of internal energy by Fig. \ref{fig4}. Neglecting correction, we can see internal energy has a maximum, while in presence of quantum corrections there is also a minimum. It may be corresponding to stable point at small radius which discussed already.

\subsection{Hyperbolic space}
In the case of hyperbolic
space $(k=-1)$ the black hole mass in the horizon given by,
\begin{equation}\label{43}
M=H=\frac{1}{a}(\frac{-\Lambda_W}{r_h})^\frac{1}{2}[-1-\Lambda_W
r_h^2]
\end{equation}
It has been
find that the magnitude of the cosmological constant increases the
black hole mass. Therefore, one can obtain black hole
entropy and temperature as follow,
\begin{equation}\label{44}
S_{0}=\frac{8\pi}{a}\sqrt{-\Lambda_W r_h},
\end{equation}
and,
\begin{equation}\label{45}
T=\frac{1}{8\pi r_h}(1-3\Lambda_W r_h^2).
\end{equation}
It is clear that the magnitude of the cosmological constant increases the black
hole temperature.
In that case Helmkoltz free energy obtained as,
\begin{eqnarray}\label{62-2}
F&=&-2\sqrt {{\frac {l}{r_h}}}\frac{1+l{r_h}^{2}}{a}\nonumber\\
&+&{\frac {\alpha}{8\pi \,r_h} \left(  -\left( 1+3\,l{r_h}^{2} \right) \ln  \left( 8\,{\frac {
\pi \sqrt {l r_h}}{a}} \right) +\frac{1}{2}(-1+3l{r_h}^{2})\right) }\nonumber\\
&-&{\frac {a\gamma\,\sqrt {l}}{32{\pi }^{2}\sqrt {r_h}} \left({\frac {1}{3l{r_h}^{2}}}+3 \right) }.
\end{eqnarray}

\begin{figure}[h!]
 \begin{center}$
 \begin{array}{cccc}
\includegraphics[width=75 mm]{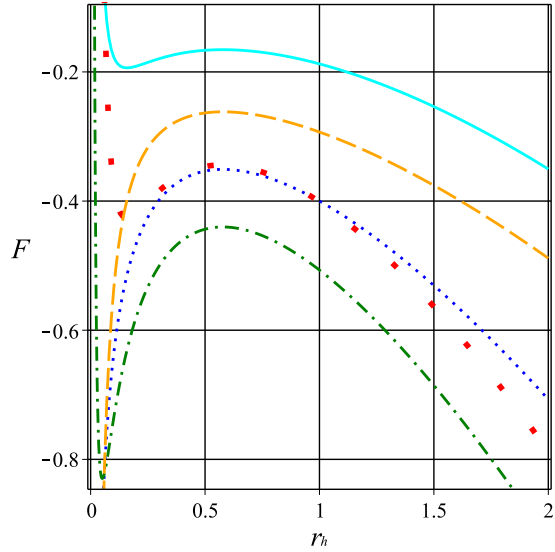}
 \end{array}$
 \end{center}
\caption{Helmholtz free energy in terms of horizon radius for the LMP solution of HL black hole in hyperbolic space with $l=1$ and $a=10$.}
 \label{fig5}
\end{figure}

As we can see from the Fig. \ref{fig5}, Helmholtz free energy of this case is completely negative with an extremum for the uncorrected entropy and two extrema  for the higher order corrected entropy.\\
One can obtain pressure as follow,
\begin{equation}\label{64-2}
P=\frac{3lr_{h}^{2}+1}{16\pi r_{h}}\left[\frac{8\pi}{a}\sqrt{lr_{h}}+\alpha\ln{\frac{8\pi\sqrt{lr_{h}}}{a}}+\frac{\gamma a}{8\pi \sqrt{lr_{h}}}\right],
\end{equation}
while the black hole volume by using the equation (\ref{17}) obtained as follow,
\begin{equation}\label{VLMPSP-2}
V=\frac{128(3lr_{h}^{2}+1)\pi^{2}l^{2}r_{h}^{2}}{64(9lr_{h}^{2}-1)\pi^{2}l^{2}r_{h}^{2}+8\pi a (lr_{h})^{\frac{3}{2}}\alpha(3lr_{h}^{2}-1)(1+2\ln{\frac{8\pi\sqrt{lr_{h}}}{a}})+3la^{2}r_{h}\gamma(lr_{h}^{2}-1)}.
\end{equation}

\begin{figure}[h!]
 \begin{center}$
 \begin{array}{cccc}
\includegraphics[width=55 mm]{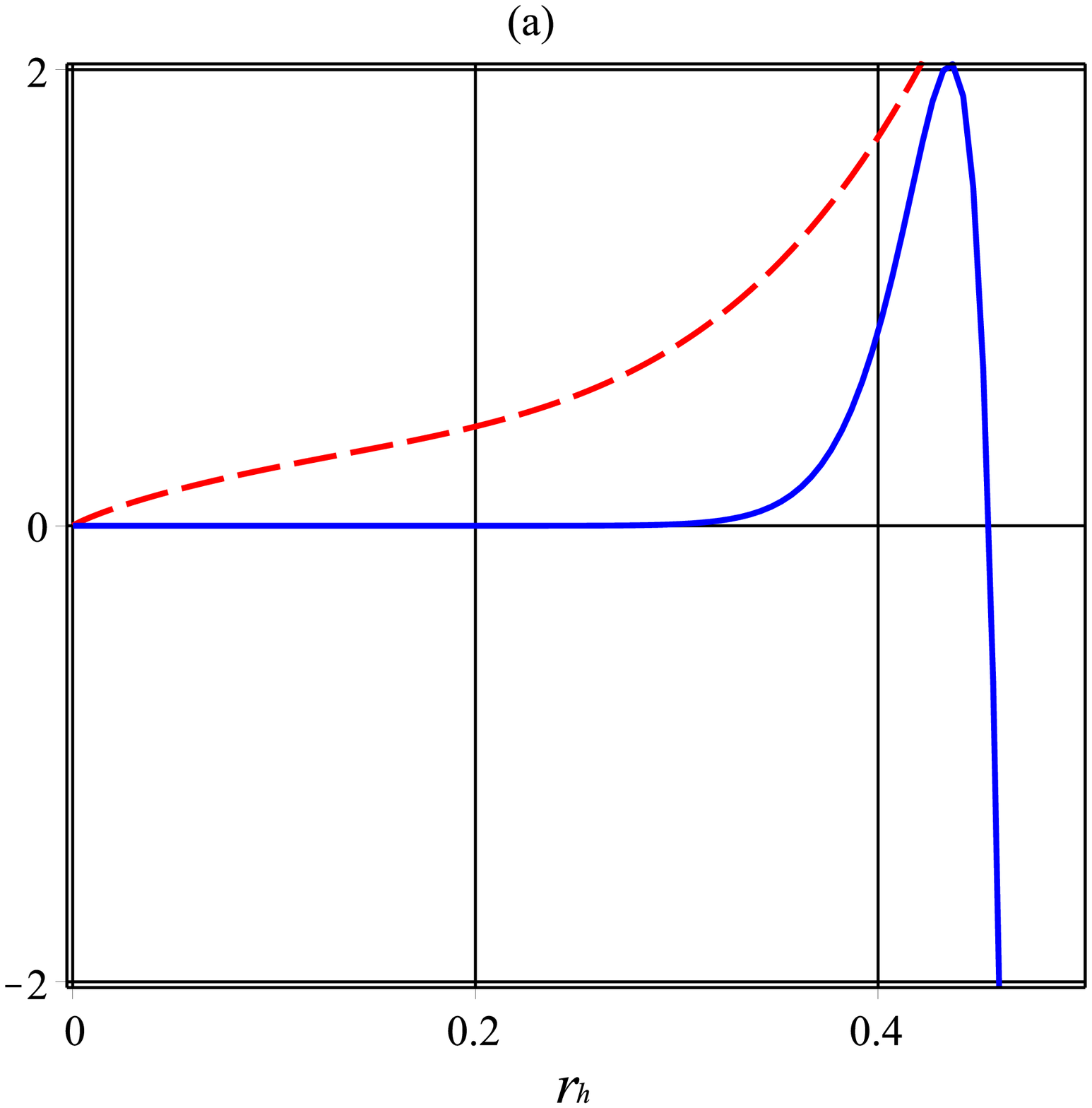}\includegraphics[width=55 mm]{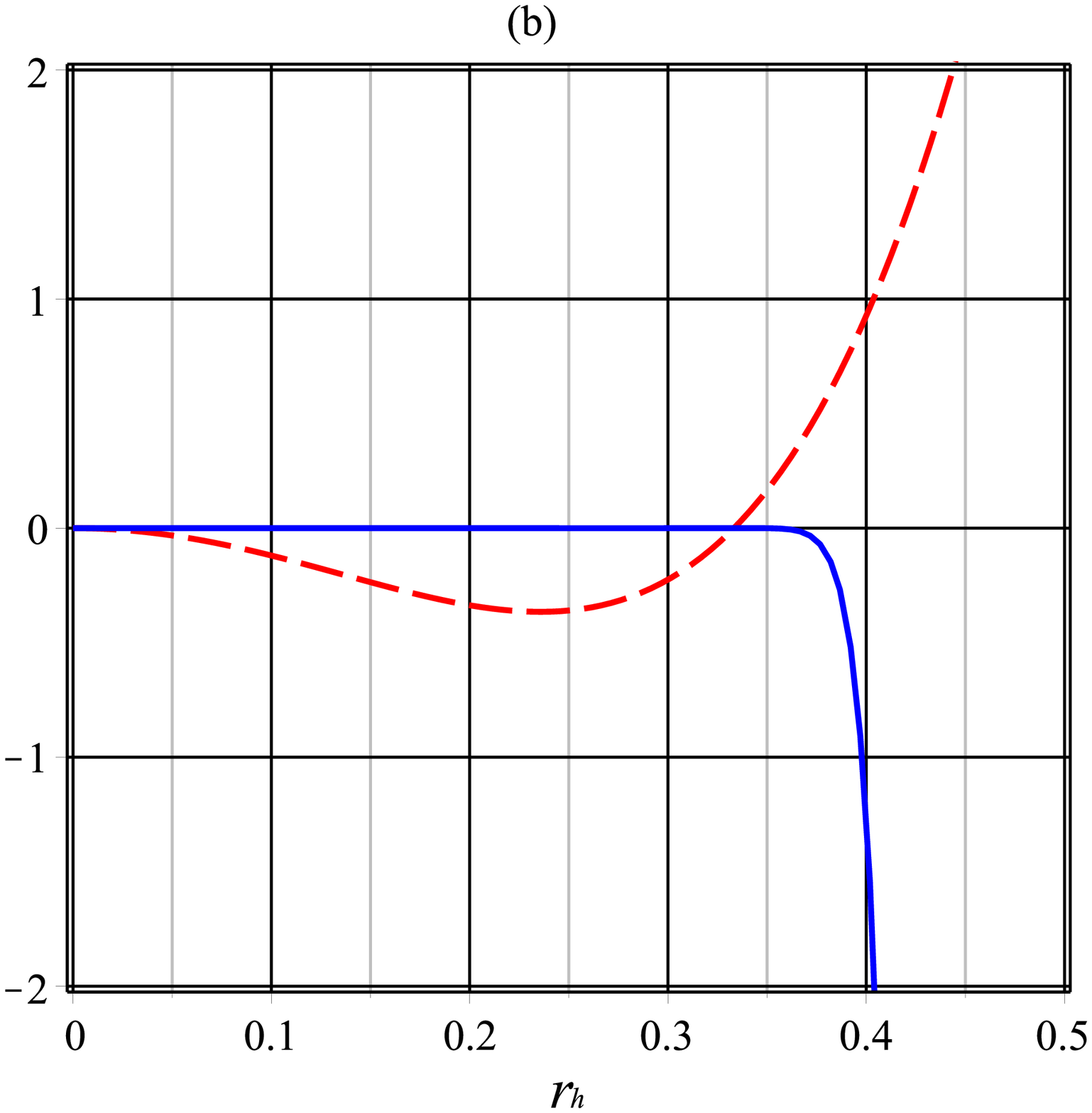}\includegraphics[width=55 mm]{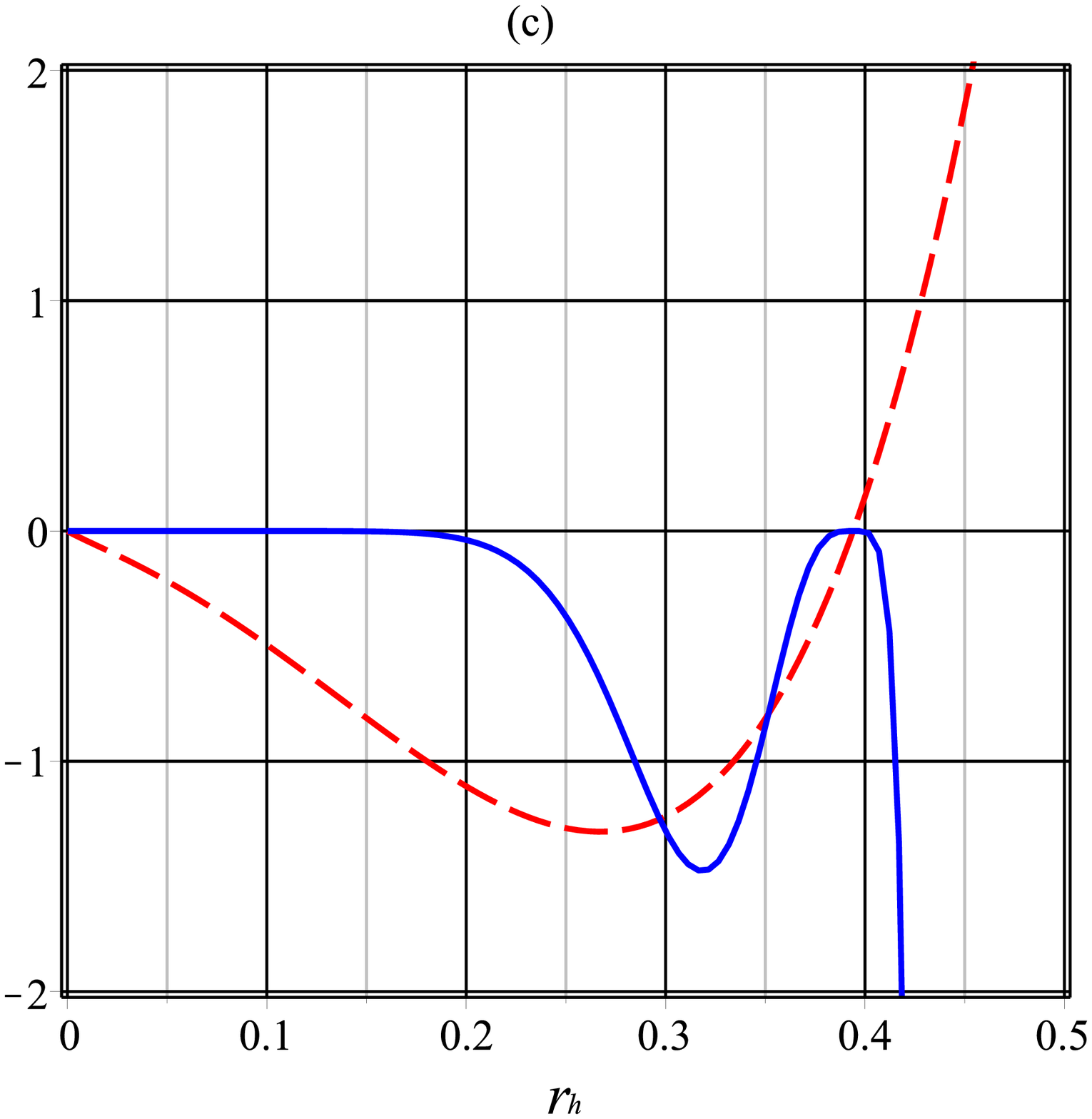}
 \end{array}$
 \end{center}
\caption{Critical points of LMP solution in hyperbolic space for $l=1$ and $a=10$: (a) $\alpha=\gamma=-1$. (b) $\alpha=\gamma=0$. (c) $\alpha=\gamma=1$. Dashed red lines represent $\frac{dP}{dV}$, while solid blue lines represent $\frac{d^{2}P}{dV^{2}}$.}
 \label{fig6}
\end{figure}

\begin{figure}[h!]
 \begin{center}$
 \begin{array}{cccc}
\includegraphics[width=75 mm]{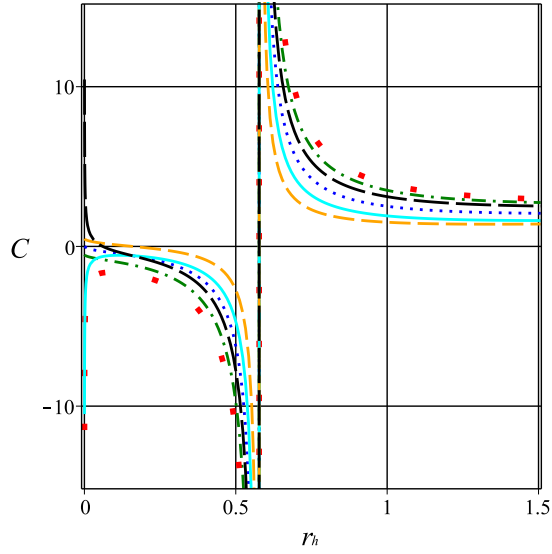}
 \end{array}$
 \end{center}
\caption{Specific heat of LMP solution in hyperbolic space in terms of the black hole horizon for $l=1$ and $a=10$.}
 \label{fig7}
\end{figure}

From the plots of Fig. \ref{fig6} we can see the uncorrected case has critical point near $r_{h}\approx0.325$ (see Fig. \ref{fig6} (b)). Also, higher order corrected case with positive coefficients have critical point near $r_{h}\approx0.4$ (see Fig. \ref{fig6} (c)). However, opposite to the spherical case, LMP solution in hyperbolic space in presence of higher order corrections with negative coefficients (Fig. \ref{fig6} (a)) don't have critical point. It means that higher order corrections affect critical points and remove them.\\
we can calculate heat capacity as,
\begin{equation}\label{29-2}
C=\frac{3lr_{h}^{2}+1}{16\pi a(lr_{h})^{\frac{3}{2}}(3lr_{h}^{2}-1)}\left(64\pi^{2}l^{2}r_{h}^{2}+8\pi a (lr_{h})^{\frac{3}{2}}\alpha-la^{2}r_{h}\gamma\right).
\end{equation}
It is clear that logarithmic correction increase value of heat capacity while the second order correction reduces its value which is similar to the previous case. In the Fig. \ref{fig7} we can see behavior of specific heat of LMP solution of HL black hole in hyperbolic space and see that there is phase transition in all cases of corrected and uncorrected entropy. It means that higher order corrections have no any effects on the phase transition.\\

\begin{figure}[h!]
 \begin{center}$
 \begin{array}{cccc}
\includegraphics[width=75 mm]{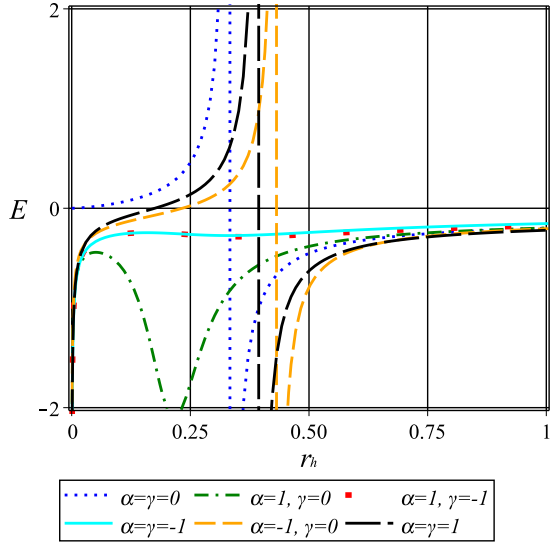}
 \end{array}$
 \end{center}
\caption{Internal energy of LMP solution in hyperbolic space in terms of the black hole horizon for $l=1$ and $a=10$.}
 \label{fig8}
\end{figure}

Finally, in the Fig. \ref{fig8} we can see behavior of the internal energy of LMP solution in hyperbolic space. We can see that in presence of higher order corrections the internal energy is completely negative (see solid cyan line of Fig. \ref{fig8}). We can see that increasing or decreasing internal energy is depend on sign of correction coefficients.

\subsection{Flat space}
This case obtained by setting $k=0$, hence the black hole mass given by,
\begin{equation}\label{22-3}
M=a^{-1}(-\Lambda_W r_h ^{2})^{\frac{3}{2}}.
\end{equation}
By using the equation (\ref{13}) one can obtain the entropy and Hawking temperature as follow,
\begin{equation}\label{23-3}
S_{0}=\frac{8\pi}{a}\sqrt{-\Lambda_W r_h},
\end{equation}
and
\begin{equation}\label{24-3}
T=-\frac{3}{8\pi}\Lambda_W r_h.
\end{equation}
It has been also finding that the magnitude of the cosmological constant increases the black hole temperature.\\
Now, we can use the equation (\ref{CS}) to obtain corrected thermodynamics due to higher order quantum corrections.\\
As before, the black hole Helmholtz free energy obtained as
\begin{equation}\label{62-3}
F=-\frac{2}{a}(l r_{h})^{\frac{3}{2}}+\frac{3}{8\pi}l r_{h}\alpha\left(\frac{1}{2}-\ln{\frac{8\pi\sqrt{l r_{h}}}{a}}\right)-\frac{3a}{32\pi^{2}}\sqrt{l r_{h}}\gamma,
\end{equation}
where we set $\Lambda_{W}=-l$ as previous subsections.\\

\begin{figure}[h!]
 \begin{center}$
 \begin{array}{cccc}
\includegraphics[width=75 mm]{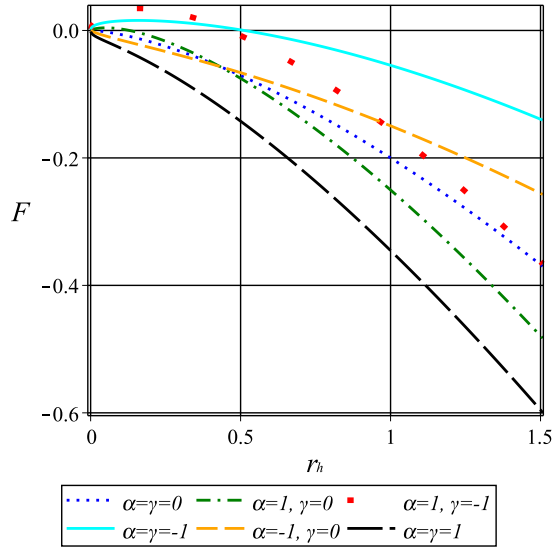}
 \end{array}$
 \end{center}
\caption{Helmholtz free energy of the LMP solution of HL black hole in flat space for $l=1$ and $a=10$.}
 \label{fig9}
\end{figure}

\begin{figure}[h!]
 \begin{center}$
 \begin{array}{cccc}
\includegraphics[width=55 mm]{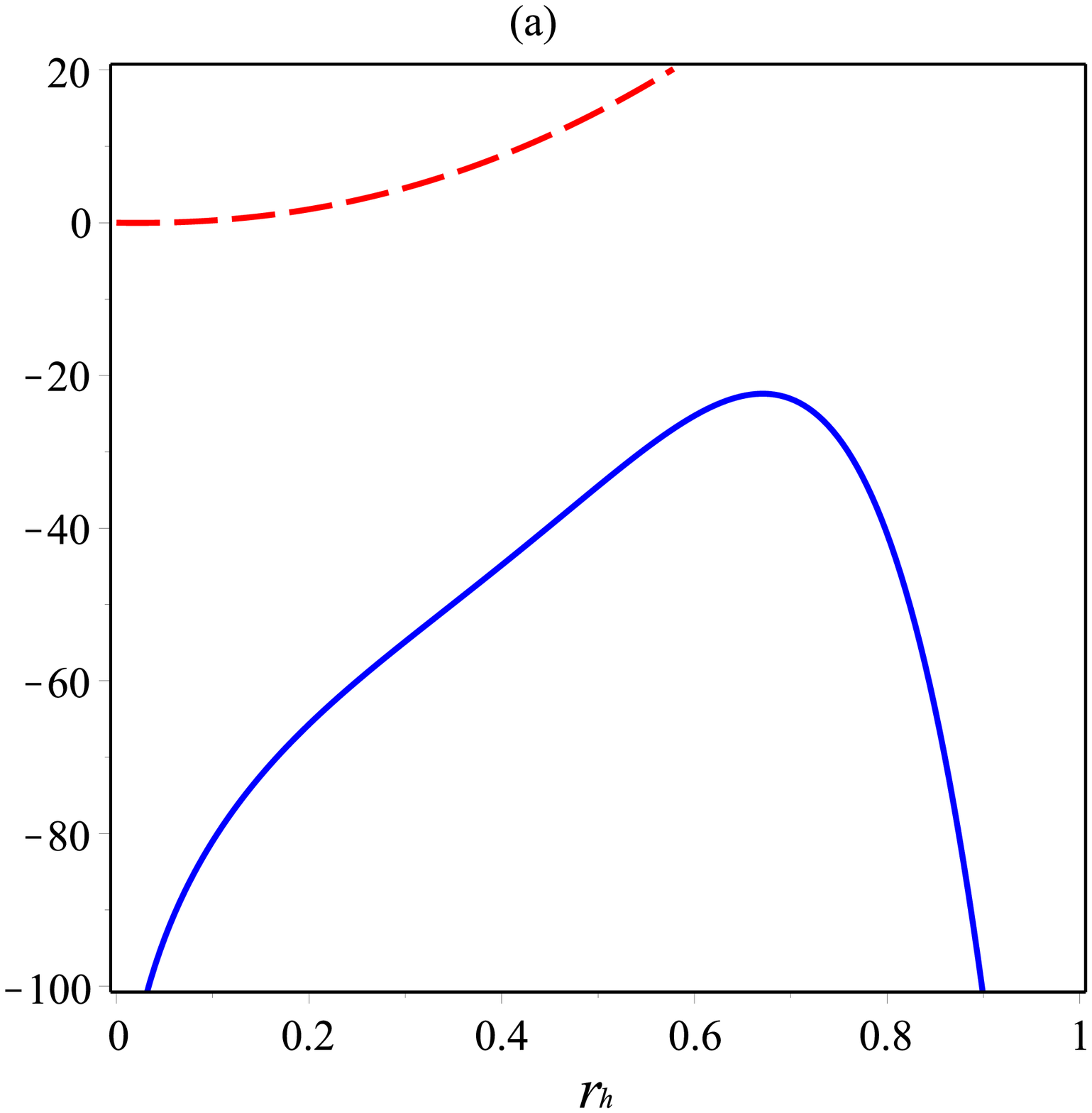}\includegraphics[width=55 mm]{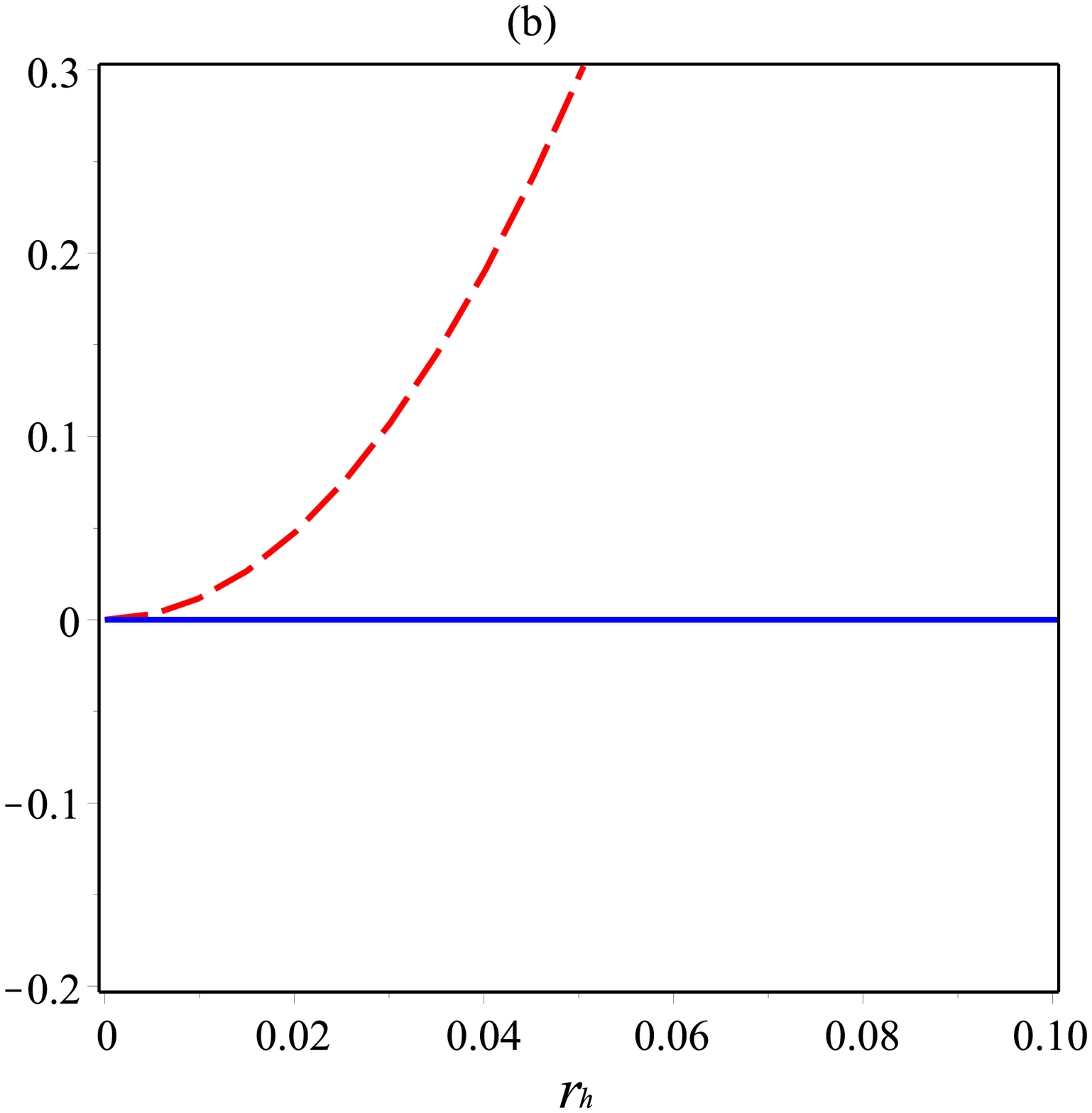}\includegraphics[width=55 mm]{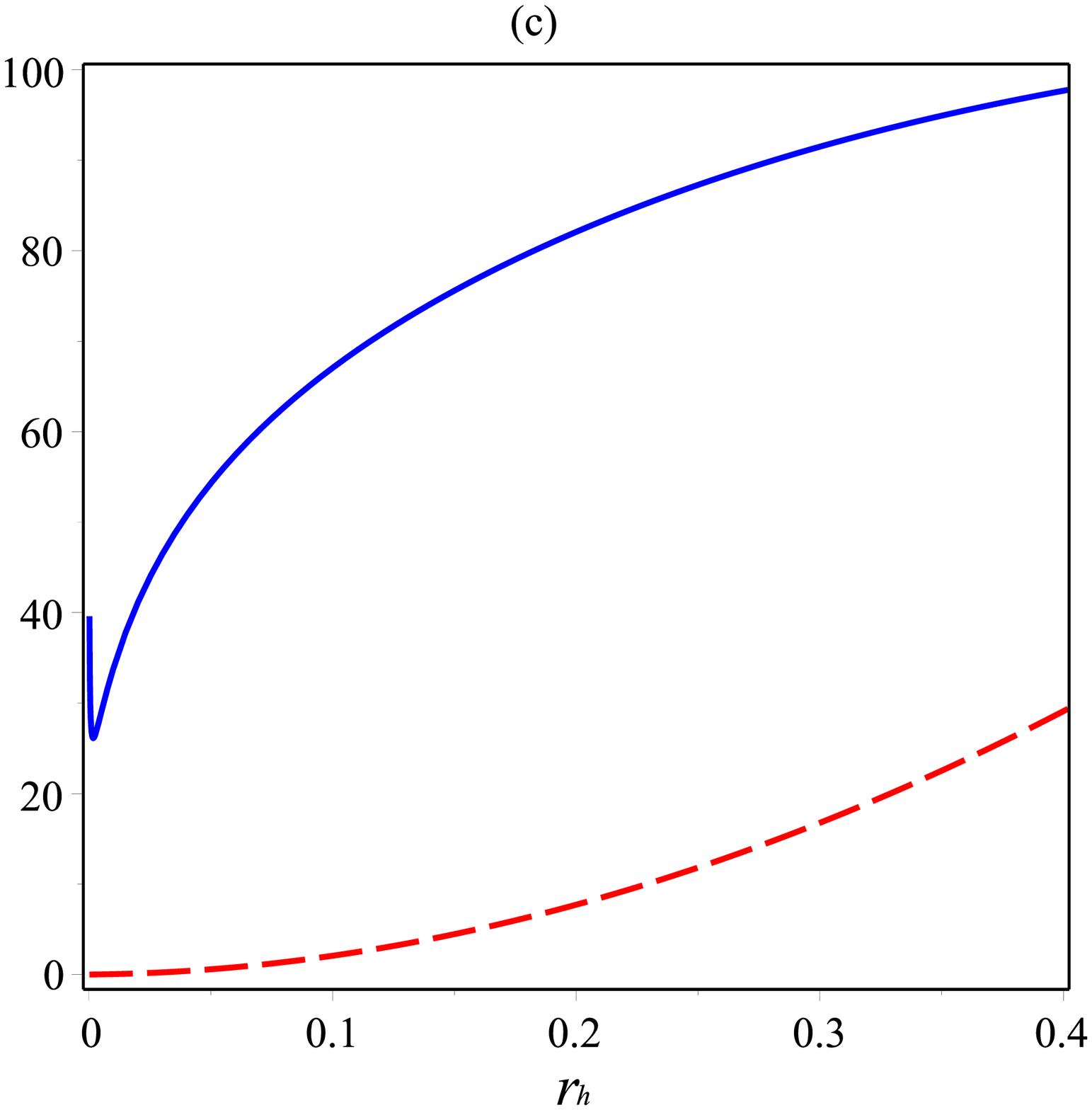}
 \end{array}$
 \end{center}
\caption{Critical points of LMP solution in flat space for $l=1$ and $a=10$: (a) $\alpha=\gamma=-1$. (b) $\alpha=\gamma=0$. (c) $\alpha=\gamma=1$. Dashed red lines represent $\frac{dP}{dV}$, while solid blue lines represent $\frac{d^{2}P}{dV^{2}}$.}
 \label{fig10}
\end{figure}

In the Fig. \ref{fig9} we can see behavior of Helmholtz free energy of the LMP solution in flat space and find important effect of quantum corrections. Neglecting thermal fluctuations, the Helmholtz free energy is completely negative. In presence of higher order corrections (solid cyan line of the Fig. \ref{fig9}) the Helmholtz free energy has a maximum in positive regions.\\
One can obtain pressure as follow,
\begin{equation}\label{64-3}
P=\frac{3lr_{h}}{16\pi}\left[\frac{8\pi}{a}\sqrt{lr_{h}}+\alpha\ln{\frac{8\pi\sqrt{lr_{h}}}{a}}+\frac{\gamma a}{8\pi \sqrt{lr_{h}}}\right].
\end{equation}
Then, the black hole volume obtained as follow,
\begin{equation}\label{VLMPSP-3}
V=\frac{128l^{2}\pi^{2}r_{h}^{2}}{192\pi^{2}l^{2}r_{h}^{2}+16\pi a (lr_{h})^{\frac{3}{2}}\alpha(\frac{1}{2}+\ln{\frac{8\pi\sqrt{lr_{h}}}{a}})+la^{2}r_{h}\gamma}.
\end{equation}
The points which satisfy condition (\ref{CriticalCondition}) illustrated by plots of the Fig. \ref{fig10}. Critical points analysis for the ordinary case ($\alpha=\gamma=0$ which given by the Figs. \ref{fig10} (b)) and corrected cases (Figs. \ref{fig10} (a) and (c)) show that there is no critical point except at $r_{h}=0$ of uncorrected case (see Figs. \ref{fig10} (b)).\\

\begin{figure}[h!]
 \begin{center}$
 \begin{array}{cccc}
\includegraphics[width=75 mm]{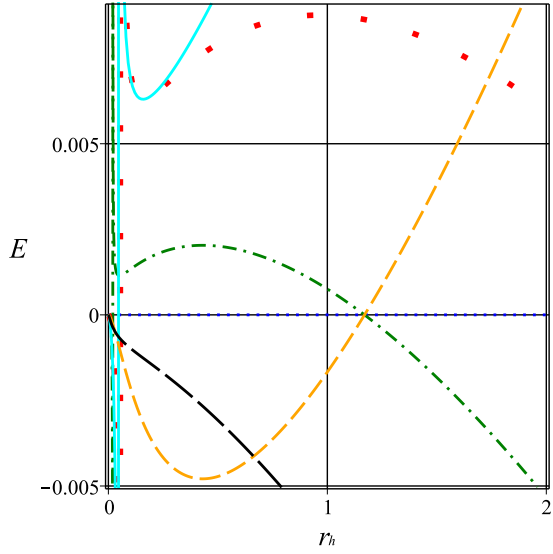}
 \end{array}$
 \end{center}
\caption{Internal energy of LMP solution in flat space in terms of the black hole horizon for $l=1$ and $a=10$.}
 \label{fig11}
\end{figure}

Heat capacity of this case given by,
\begin{equation}\label{29-3}
C=r_{h}\left[\frac{4\pi}{a}\sqrt{\frac{l}{r_{h}}}+\frac{\alpha}{2r_{h}}-\frac{a\gamma}{16\pi l r_{h}^{\frac{3}{2}}}\right].
\end{equation}
It is clear that logarithmic correction increases value of heat capacity while the second order correction reduces its value.\\
In absence of higher order corrections, the heat capacity is positive. Logarithmic correction reduces value of heat capacity to make it negative for the small black hole, when $r_{h}$ is large, the effect of thermal fluctuation is negligible. Higher order correction with positive coefficient also decrease value of heat capacity hence affect black hole stability. In the case of $\alpha=\gamma=-1$ the heat capacity has a minimum.\\
Finally, we give graphical analysis of internal energy by Fig. \ref{fig11}. Effect of quantum corrections are important which may reduce or increase value of internal energy as illustrated by the Fig. \ref{fig11}.

\section{Kehagius-Sfetsos solution}
\begin{figure}[h!]
 \begin{center}$
 \begin{array}{cccc}
\includegraphics[width=60 mm]{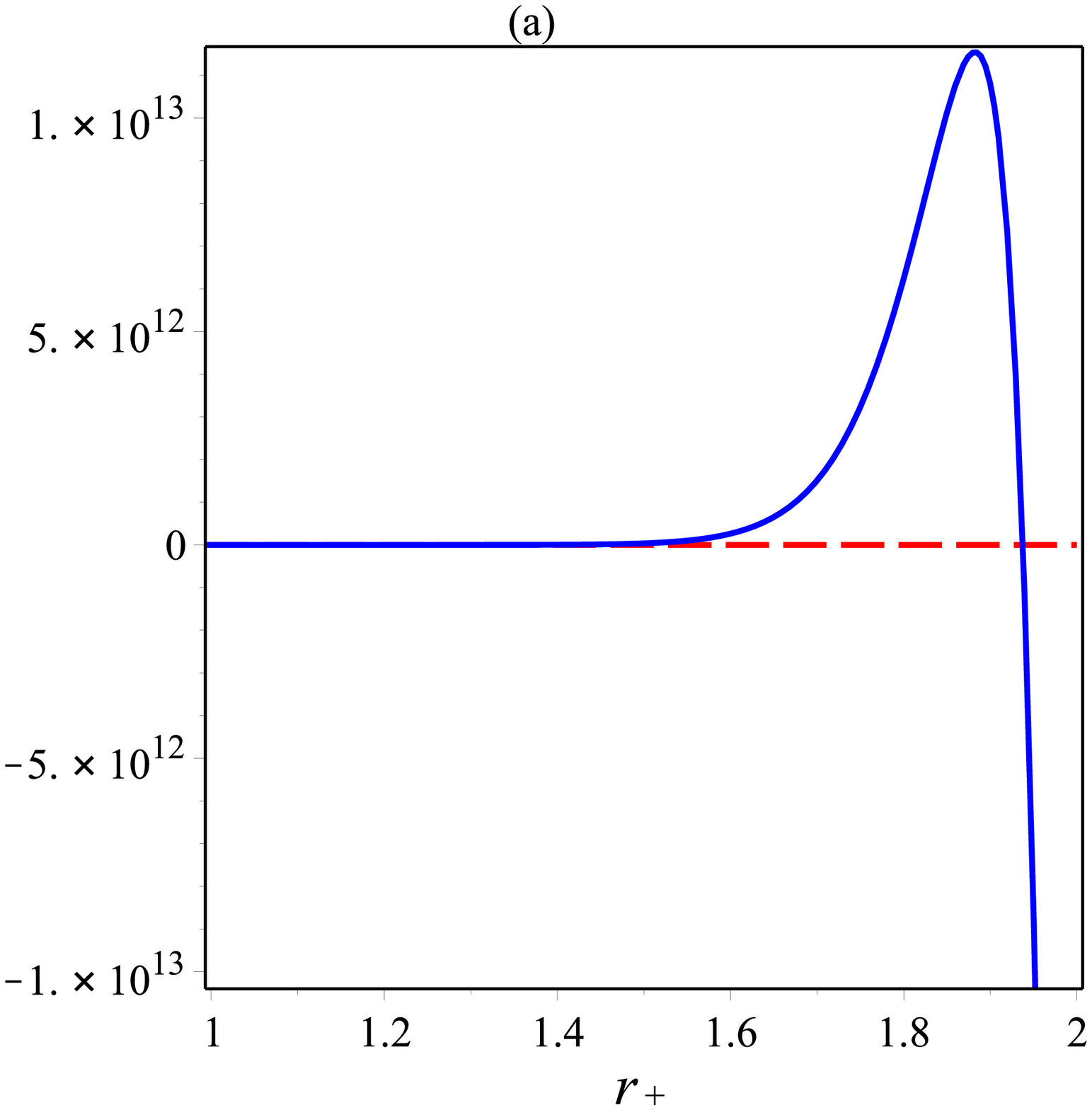}\includegraphics[width=60 mm]{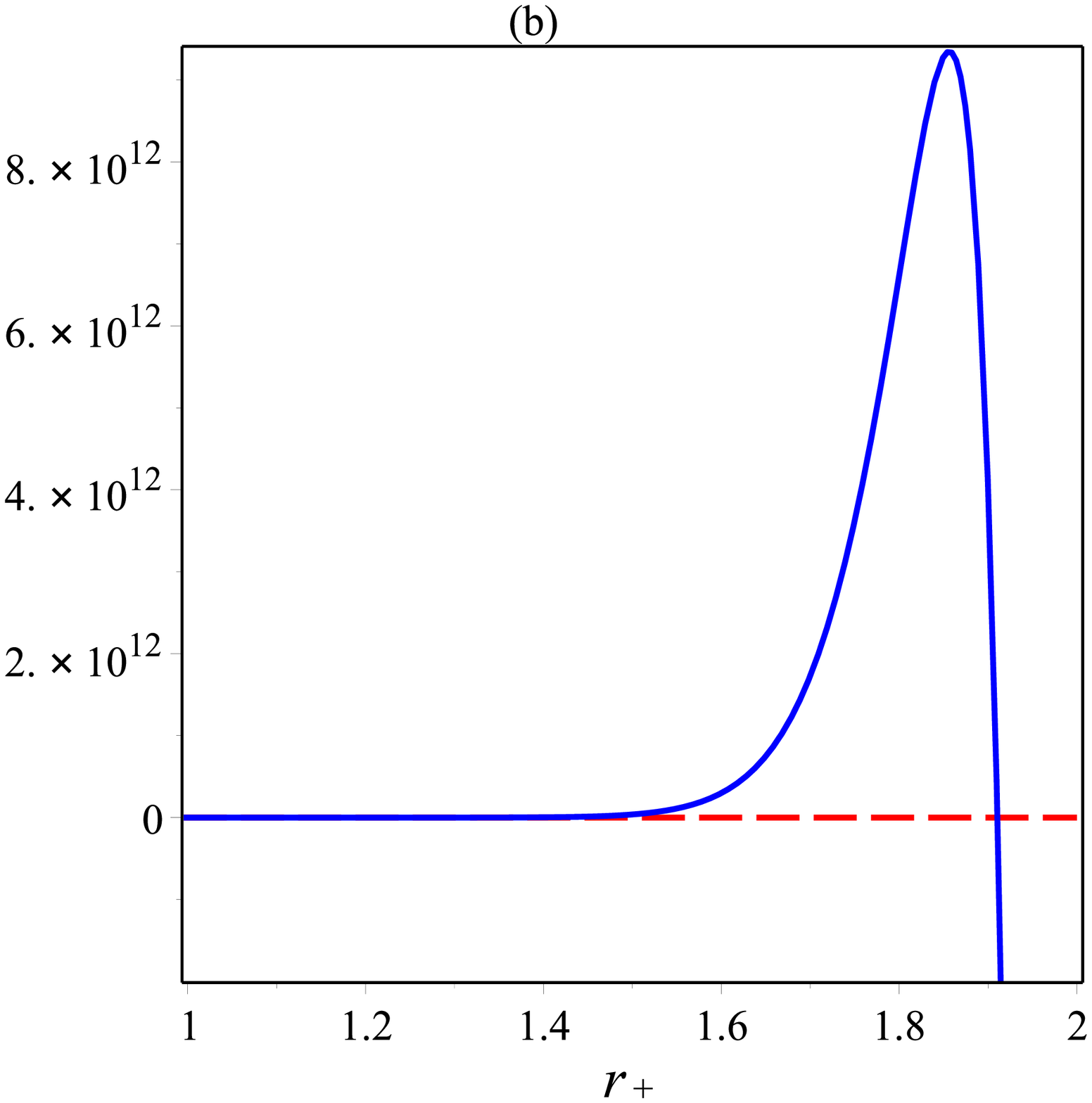}
 \end{array}$
 \end{center}
\caption{Critical points of KS solution $\omega=1$ and $k=1$: (a) $\alpha=\gamma=-1$. (b) $\alpha=\gamma=0$. Dashed red lines represent $\frac{dP}{dV}$, while solid blue lines represent $\frac{d^{2}P}{dV^{2}}$.}
 \label{fig12}
\end{figure}

In this case, by using the equation (\ref{8}) one can obtain inner and outer horizons as,
\begin{equation}\label{h}
r_{\pm}=\frac{2\omega M\pm\sqrt{4\omega^{2}M^{2}-2\omega k^{3}}}{2k\omega}.
\end{equation}
Hence, the black hole mass given by,
\begin{equation}\label{M-KS}
M=\frac{k(2\omega r_{+}^{2}+k)}{4\omega r_{+}}.
\end{equation}
Also, black hole temperature obtained as,
\begin{equation}\label{T-KS}
T=\frac{\omega r_{+}^{3}(\sqrt{1+\frac{4M}{\omega r_{+}^{3}}}-1)-M}{2\pi r_{+}^{2}\sqrt{1+\frac{4M}{\omega r_{+}^{3}}}}.
\end{equation}
Inserting $M$ from (\ref{M-KS}) into the temperature (\ref{T-KS}) one can obtain,
\begin{equation}\label{T-KS-2}
T=\frac{k(2\omega r_{+}^{2}-k)}{8\pi r_{+}(\omega r_{+}^{2}+k)}.
\end{equation}
Then, we can obtain entropy as follow,
\begin{equation}\label{s-KS}
S_{0}=\pi r_{+}^{2}+\frac{2\pi k}{\omega}\ln{r_{+}}.
\end{equation}
Therefore, we can obtain corrected entropy due to higher order corrections, and calculate thermodynamics quantities.\\
Corrected pressure calculated as,
\begin{equation}\label{P-KS}
P=\frac{k(\omega r_{+}^{2}-\frac{k}{2})}{8\pi r_{+}(\omega r_{+}^{2}+k)}\left[\pi r_{+}^{2}+\frac{2\pi k}{\omega}\ln{r_{+}}
+\alpha\ln{\frac{\pi}{\omega}(\omega r_{+}^{2}+2k\ln{r_{+}})}+\gamma\frac{\omega}{\pi(\omega r_{+}^{2}+2k\ln{r_{+}})}\right].
\end{equation}
Then, we can obtain black hole volume and find that logarithmic correction decreases it while the second order correction increases the black hole volume. Also, as illustrated by plots of the Fig. \ref{fig12}, higher order corrections do not affect the critical point. In absence of quantum correction there is a critical point as well as in presence of higher order corrections.

\begin{figure}[h!]
 \begin{center}$
 \begin{array}{cccc}
\includegraphics[width=65 mm]{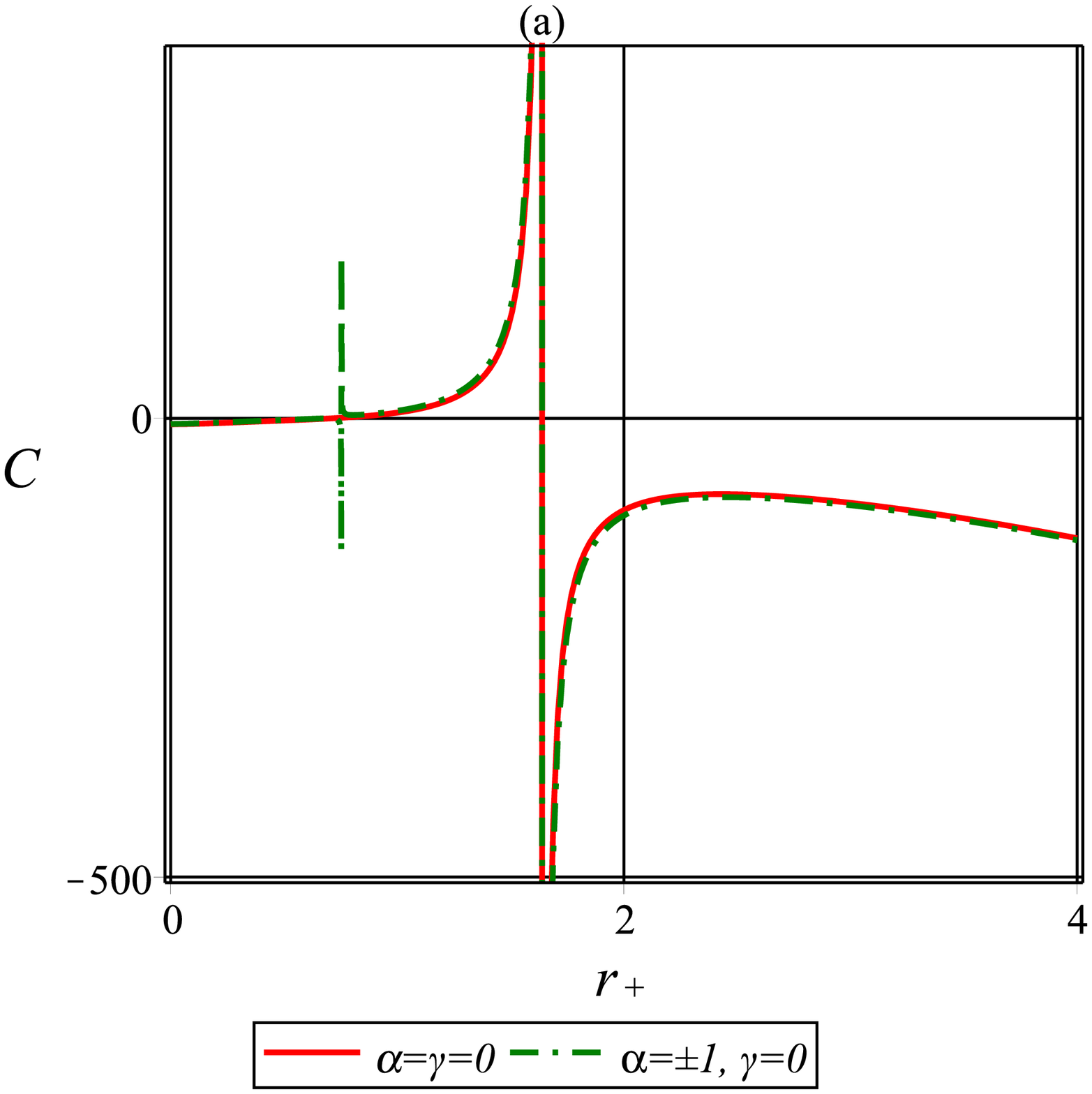}\includegraphics[width=65 mm]{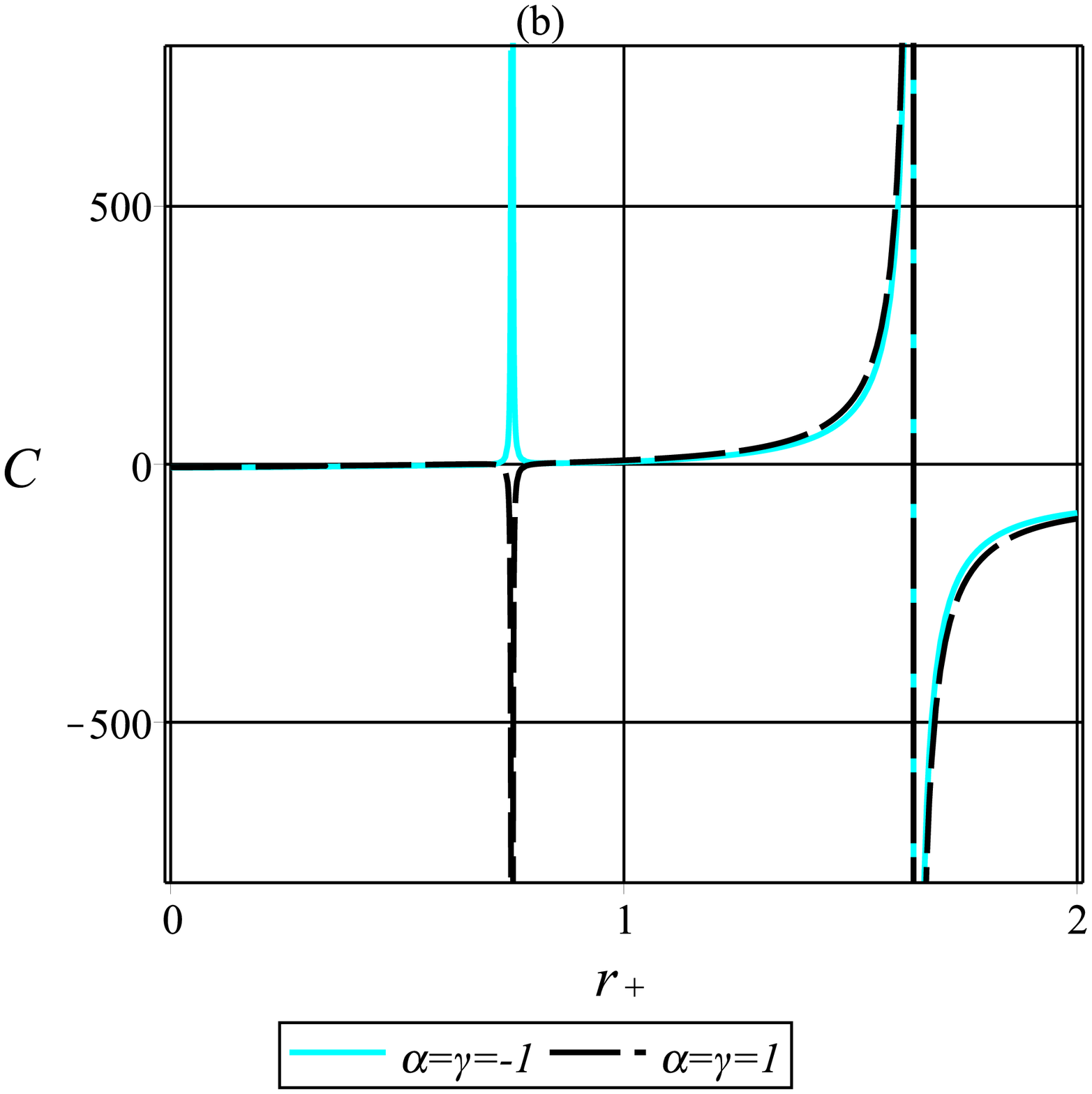}
 \end{array}$
 \end{center}
\caption{Specific heat of KS solution in terms of the black hole horizon for $\omega=1$ and $k=1$.}
 \label{fig13}
\end{figure}

Finally, we can calculate heat capacity as,
\begin{eqnarray}\label{29-4}
C&=&\frac{2(2\omega r_{+}^{2}-k)(\omega r_{+}^{2}+k)^2}{\pi\omega(-2\omega^{2}r_{+}^{4}+5k\omega r^{2}+k^{2})(0.5\omega r_{+}^{2}+k\ln{r})^{2}}\nonumber\\
&\times&\left((k\pi\ln{r_{+}})^{2}+\omega k \pi^{2}r_{+}^{2}\ln{r_{+}}+\frac{\omega^{2}\pi^{2}r_{+}^{4}}{4}+\alpha(\frac{\pi\omega^{2}r_{+}^{2}}{4}+\frac{k\pi\omega}{2}\ln{r_{+}})
-\frac{\omega^{2}}{4}\gamma\right).
\end{eqnarray}
In the Fig. \ref{fig13} we can see that quantum correction has no any important effect of the first order phase transition and black hole stability. However, presence of the logarithmic correction (see dash dot green line of the Fig. \ref{fig13}(a)) make another asymptotic region at small radius which may be interpreted as the second order phase transition.

\section{Conclusion}
In this paper, we considered Ho\v{r}ava-Lifshitz black hole to study its thermodynamics in presence of quantum correction by consideration of thermal fluctuations. We considered two special case of Kehagius-Sfetsos and Lu-Mei-Pop solutions to study critical points and thermodynamics stability separately. In the case of KS solution, we consider general case and found that logarithmic correction may yield to the second order phase transition. Higher order correction with positive coefficient may yield to some instabilities at small event horizon radius (see Fig. \ref{fig13}(b)). We found that there are also critical points in presence of higher order correction as well as in absence of quantum corrections.\\
In the case of LMP solution, we considered three cases of flat, spherical and hyperbolic spaces and studied modified thermodynamics separately. Both analytical and numerical study on the thermodynamics quantities under effects of quantum corrections show that quantum corrections are important when the black hole size reduced due to the Hawking radiation. In all cases we found that Helmholtz free energy is decreasing function of correction
parameters. Then, we obtained pressure and volume to study critical points. In the case of LMP solution with spherical space we found that higher order corrections are important to have critical point. Fig. \ref{fig2} (a) shows that only higher order corrections with negative coefficients yields to critical point which is opposite to hyperbolic space. It means that LMP solution in hyperbolic space have critical point for higher order corrections with positive coefficients. In flat space there is no critical point and quantum corrections have no any effects on critical points.\\
We study black hole stability by using heat capacity. We found that first order correction increases value of heat capacity while the second order
correction reduces its value. In the case of spherical space, we found some stable points at very small radius due to higher order corrections and black hole is stable at quantum scales. In the case of hyperbolic space, we found that
there is phase transition in all cases of corrected and uncorrected entropy. It means that
higher order corrections have no any effects on the black hole phase transition. In the case of flat space heat capacity is completely positive without higher order corrections. Logarithmic correction
reduces value of heat capacity to make it negative for the small black hole, when event horizon radius grows up, then the effect of thermal fluctuation is negligible. Higher order correction with positive
coefficient also decrease value of the heat capacity, hence change the black hole stability.\\
In this paper, we considered only presence or absence of corrections by choosing $\alpha$ and $\gamma$ as $0$ or $\pm1$. It is interesting to calculate exact values of these parameters which may be yields to experimental setup to test quantum gravity. It may be aim of our future study.\\
In summary, this paper is extension of previous work to include higher order correction to LMP solution, while logarithmic and higher order corrections of KS solution of HL black holes. It is interesting to consider higher order correction of a new regular black hole \cite{reg}, Myerse-Perry black holes \cite{MP}, rotating charged hairy black hole \cite{hair1, hair2}, or five dimensional AdS black hole at N=2 supergravity \cite{A1, A2}.\\
Recently, Thermodynamic of a black hole surrounded by perfect fluid in Rastall theory have been studied \cite{ref8}. Now, it is interesting to obtain thermal fluctuation effects on such kind of solutions.\\\\\\

{\bf Acknowledgments}: Author would like to thanks Iran Science Elites Federation.

\end{document}